\documentclass[twocolumn]{aastex62}

\usepackage{bm}
\usepackage{amsmath}

\accepted{September 17, 2019}

\shorttitle{Unveiling Dust Aggregate Structure in Protoplanetary Disks}
\shortauthors{Tazaki et al.}

\begin{document}

\title{Unveiling Dust Aggregate Structure in Protoplanetary Disks by Millimeter-wave Scattering Polarization}

\correspondingauthor{Ryo Tazaki}
\email{rtazaki@astr.tohoku.ac.jp}

\author[0000-0003-1451-6836]{Ryo Tazaki}
\affil{Astronomical Institute, Graduate School of Science
Tohoku University, 6-3 Aramaki, Aoba-ku, Sendai 980-8578, Japan}

\author{Hidekazu Tanaka}
\affil{Astronomical Institute, Graduate School of Science
Tohoku University, 6-3 Aramaki, Aoba-ku, Sendai 980-8578, Japan}

\author{Akimasa Kataoka}
\affil{National Astronomical Observatory of Japan, Mitaka, Tokyo 181-8588, Japan}

\author{Satoshi Okuzumi}
\affil{Department of Earth and Planetary Sciences, Tokyo Institute of Technology, Meguro-ku, Tokyo, 152-8551, Japan}

\author{Takayuki Muto}
\affil{Division of Liberal Arts, Kogakuin University, 1-24-2 Nishi-Shinjuku, Shinjuku-ku, Tokyo 163-8677, Japan}

\nocollaboration



\begin{abstract}
Dust coagulation in a protoplanetary disk is the first step of planetesimal formation. 
However, the pathway from dust aggregates to planetesimals remains unclear. 
Both numerical simulations and laboratory experiments have suggested the importance of dust structure in planetesimal formation, but it is not well constrained by observations.
We study how the dust structure and porosity alters polarimetric images at millimeter wavelength by performing 3D radiative transfer simulations. Aggregates with different porosity and fractal dimension are considered. As a result, we find that dust aggregates with lower porosity and/or higher fractal dimension are favorable to explain the observed millimeter-wave scattering polarization of disks. Although we cannot rule out the presence of aggregates with extremely high porosity, a population of dust particles with relatively compact structure is at least necessary to explain polarized-scattered waves.
In addition, we show that particles with moderate porosity show weak wavelength dependence of scattering polarization, indicating that multi-wavelength polarimetry is useful to constrain dust porosity. Finally, we discuss implications for dust evolution and planetesimal formation in disks. 
\end{abstract}




\section{Introduction}
%
%
Planetesimal formation, or the first step of planet formation, begins with coagulation of micron-sized dust particles in protoplanetary disks. 
Initially, micron-sized dust particles coagulate to form fluffy dust aggregates whose fractal dimension is about 2 \citep{Weidenschilling93, Ossenkopf93, Wurm98, Kempf99, Krause04}. However, it is still a matter of debate how fluffy dust aggregates grow to form planetesimals in disks. One possibility is that they can maintain their fluffy structure during growth, and the volume filling factor becomes as low as $10^{-4}$.
Then, coagulation and subsequent compaction of the aggregates form planetesimals \citep{Okuzumi12, Kataoka13}. Another possibility is that they are compressed at some initial moment during growth, and the volume filling factor is increased to higher than $0.01$ \citep{Zsom10, Lorek18}. These relatively compact dust aggregates are thought to evolve into planetesimals via the gravitational collapse of dust clouds led by the streaming instability \citep{Youdin05, Johansen07, Bai10a, Bai10b, Drazkowska14} or by direct coagulation with mass transfer \citep{Windmark12a, Windmark12b}. Therefore, constraining dust porosity or structure by observations seems to be a very helpful way to discriminate which planetesimal formation process dominates in disks. 

%
%
Recently, the Atacama Large Millimeter/submillimeter Array (ALMA) opened a new window to observe sub-millimeter-wave polarization of protoplanetary disks \citep{Kataoka16b, Kataoka17, Stephens17, Cox18, Sadavoy18, Hull18, Lee18, Alves18, Bacciotti18, Girart18, Ohashi18, Dent19, Takahashi19, Harrison19}. 
Several mechanisms to explain polarized millimeter-wave radiation have been proposed: scattering \citep{Kataoka15, Pohl16, Yang16a, Yang16b, Yang17, Okuzumi19}, grain alignment with magnetic field \citep{Cho07, Bertrang17}, alignment with the radiation field \citep{Lazarian07a, Tazaki17}, the gold alignment \citep{Yang19}, and mechanical alignment \citep{Lazarian07b, Kataoka19}. 

One possible origin is a scattering scenario. In the scattering model, polarized millimeter-waves are produced by scattering of thermal emission from ambient dust particles. Observed polarization patterns and polarization fractions are consistent with the predictions of the scattering model \citep{Kataoka16b, Stephens17, Hull18, Lee18, Bacciotti18, Girart18, Ohashi18, Dent19, Harrison19}. In addition, the scattering model may also explain the wavelength dependencies of the observed polarization fraction, such as for HL Tau disk \citep{Stephens17} and for DG Tau disk \citep{Bacciotti18, Harrison19}.

Interestingly, if scattering occurs in disks, it can constrain the maximum dust particle radius \citep{Kataoka15}. Previous studies assume solid spherical particles, and the porosity or structure of dust particles are ignored for the  sake of simplicity. However, the light scattering process sensitively depends on dust particle size and structure \citep{Kimura06, Shen08, Shen09, Tazaki16, Tazaki18, Ysard18}. Thus, current constraints on dust properties could be affected if dust porosity and structure are taken into account. The role of dust porosity and structure on millimeter-wave scattering polarization is not yet clarified. Therefore, this paper studies how these dust particle properties affect millimeter-wave-scattering polarization.

%
%
This paper is organized as follows. 
Section \ref{sec:model} describes our dust particle models and methods. 
Section \ref{sec:optprop} summarizes optical properties of dust particles. 
In Section \ref{sec:result}, using the optical properties obtained in Section \ref{sec:optprop}, we perform radiative transfer simulations of disks, and the results are presented. 
By comparing our results with polarimetric observations by ALMA, we discuss implications for planetesimal formation in Section \ref{sec:disc}. Section \ref{sec:sum} presents a summary.

\section{Dust Model and Method} \label{sec:model}

\subsection{Dust particle structure during growth}
Prior to presenting our dust particle models in Section \ref{sec:duststrc}, we briefly review the expected dust structure in protoplanetary disks.

The structure of dust aggregates depends on how they coagulate. 
In disks, coagulation begins with low-collision velocity: no restructuring of dust aggregates occurs upon impact at this stage. Two limiting cases for aggregation are often used: Ballistic Cluster-Cluster Aggregation (BCCA) and Ballistic Particle-Cluster Aggregation (BPCA). Aggregates consist of unit particles (monomer) of the same radius for the sake of simplicity.

BCCA is successive collisions between similar-sized aggregates. This type of coagulation is expected to occur if the aggregate-size distribution is narrow. Initial dust coagulation in disks is thought to take place in this manner \citep{Ormel07, Okuzumi09, Okuzumi12}. 
Aggregates formed by collisions of similar-sized aggregates tend to show fractal dimension $1.4\lesssim d_f\lesssim2$ \citep{Weidenschilling93, Ossenkopf93, Wurm98, Kempf99, Krause04}. Since the volume filling factor $f$ and radius of an aggregate $a$ obey $f\propto a^{d_f-3}$, their volume filling factor decreases with increasing radius. For $100~\mu$m-sized BCCA aggregates consisting $0.1~\mu$m monomers, the volume filling factor is $f=3\times10^{-4}$; thus, they are extremely fluffy.

BPCA is successive collisions of a monomer and an aggregate. 
This type of collision is realized when small and large aggregates coexist in disks, such as due to fragmentation of large aggregates. It has been suggested that fragmentation of dust aggregates seems to be necessary to explain disk infrared observations \citep{Dullemond05}.
The BPCA aggregates tend to have $d_f\approx3$, and hence, their volume filling factor is constant and is about $f=0.15$ \citep{Kozasa92}.
\citet{Dominik16} studied hierarchical coagulation (an aggregate of aggregates), where an aggregate consisting of some monomers is used as a projectile in BPCA instead of using a monomer particle. Hierarchical coagulation produces an approximately 10-times lower filling factor $f\approx0.01$, yet still shows $d_f\approx3$ at the core of the aggregate.

%
%
In disks, dust structure might also be affected by some compaction events.
For high-speed collisions such that the impact energy is high enough to rearrange aggregate structure, but still insufficient to disrupt an aggregate, collisional compaction occurs \citep{Dominik97, Blum00, Wada07, Wada08, Wada09, Paszun09}.
\citet{Wada08} showed that the fractal dimension of BCCA aggregates can be increased by a factor of up to $2.5$ upon such an impact. Laboratory measurements have often observed bouncing of two colliding dust aggregates \citep{Blum93, Weidling12, Kothe13, Brisset17}. \citet{Weidling09} suggested that sequential bouncing collisions gradually increase the volume filling factor up to $f\approx0.36$. However, the conditions for bouncing collisions are still being debated \citep{Wada11, Seizinger13a, Kothe13}.
\citet{Kataoka13a, Kataoka13} proposed gas and gravitational compaction of dust aggregates which increases the volume filling factor.
Dust structure under these compressions is characterized by a bi-fractal structure, where $d_f=3$ on the larger scale and $2$ on smaller scale. 

To summarize, dust aggregates in disks might be categorized into two limiting groups: compact dust aggregates showing fractal dimension $d_f\approx3$ with $0.01\lesssim f \lesssim0.4$ and fluffy dust aggregates with $d_f\approx2$ with $f\ll1$. 

\subsection{Dust composition} \label{sec:dustcomp}
At (sub-)millimeter wavelengths, disk emission comes mostly from the outer cold regions where water ice condenses onto dust particles.
Thus, we assume that a dust particle is a mixture of silicate, water ice, troilite, and amorphous carbon. Although silicate, troilite, and amorphous carbon are probably separate grain populations, we assume that these components are mixed into a single dust particle for the sake of simplicity.
The mass fraction of each component is determined by a recipe described in \citet{Min11}, where we adopt the carbon partition parameter $w=0.5$. The derived mass fractions (material density) of silicate, water ice, troilite, and amorphous carbon are 32\% (3.30 g cm$^{-3}$), 45\% (0.92 g cm$^{-3}$), 10\% (4.83 g cm$^{-3}$), and 13\% (1.80 g cm$^{-3}$), respectively. 
Water ice dominates the mass and volume of dust particles. 
The resulting mean material density is 1.48 g cm$^{-3}$. Optical constants of silicate, water ice, troilite, and amorphous carbon are taken from \citet{Draine03b}, \citet{Warren08}, \citet{Henning96}, and \citet{Zubko96}, respectively. 
Their optical constants are mixed by using the Bruggeman mixing rule \citep{Bruggeman35}. At $\lambda=1$ mm, the mixed refractive index, $m$, is $m=2.6+0.074i$.

\subsection{Dust particle models adopted in this study} \label{sec:duststrc}
\begin{figure}
\begin{center}
\includegraphics[height=8.0cm,keepaspectratio]{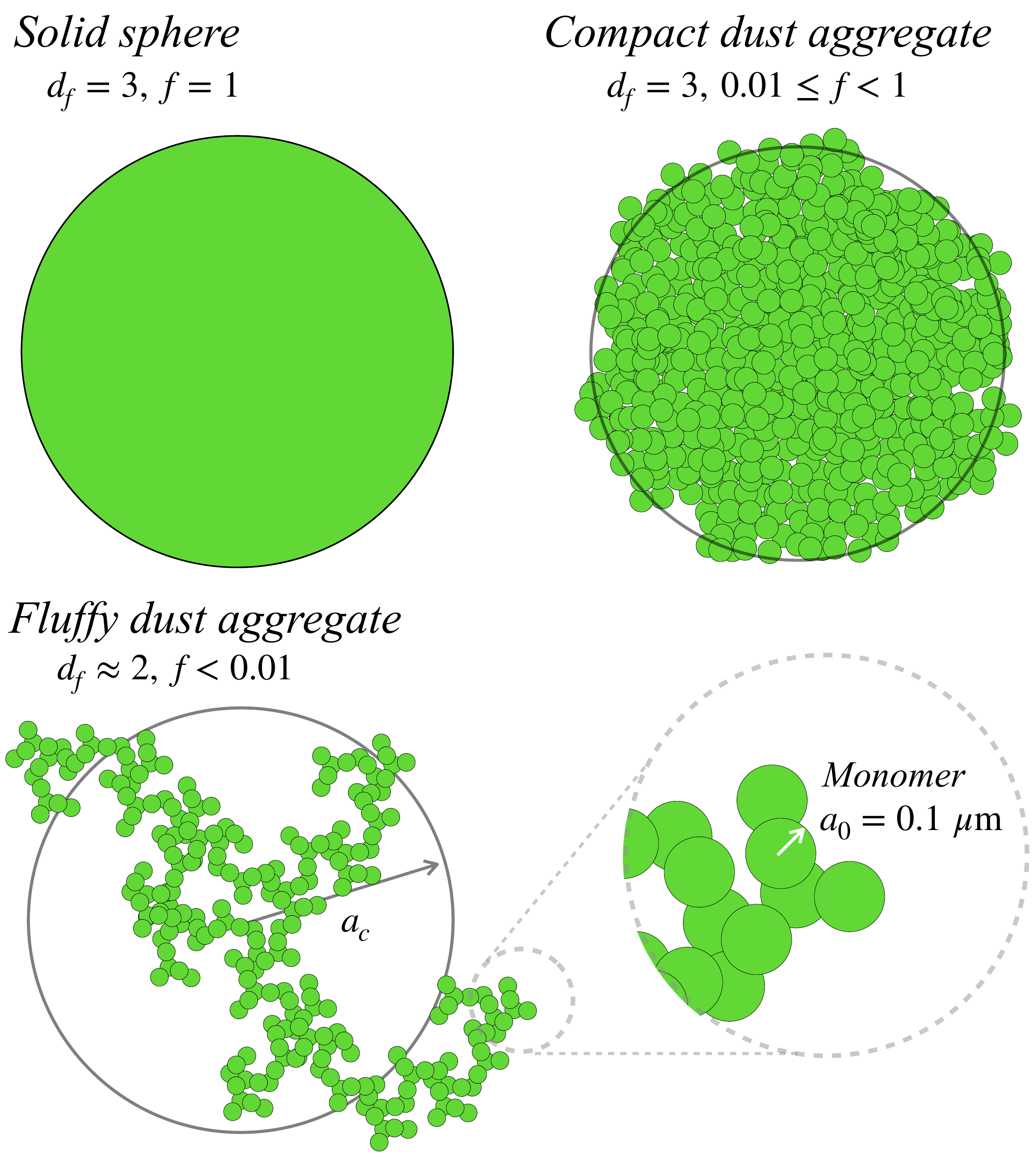}
\caption{Cartoon illustrating dust particle models used in this paper. 
Dust models are characterized by the volume filling factor $f$ and fractal dimension $d_f$. Solid spheres have $d_f=3$ and $f=1$, whereas compact dust aggregates have $d_f=3$ and $f<1$.
We also consider fluffy dust aggregates whose fractal dimension is $d_f\approx2$ consisting of unit particles (monomers) of radius $a_0=0.1~\mu$m. The volume filling factor of fluffy aggregates is size dependent and it is typically much smaller than 0.01.}
\label{fig:dustmodel}
\end{center}
\end{figure}

We consider three types of dust structure as illustrated in Figure \ref{fig:dustmodel}.
Basic properties of each dust model are as follows. 
\begin{itemize}
\item {\it Solid spheres}: Solid spheres have homogeneous structure, and the filling factor is unity. Since solid spheres are commonly used in previous studies, it is valuable to compare how their optical properties are different from those of lower density dust particles resulting from structure and porosity. The important parameter of this model is its radius.

\item {\it Fluffy dust aggregates}: We define fluffy dust aggregates as aggregates with $d_f\approx2$. 
We use the characteristic radius $a_c$ to describe the aggregate radius \citep{Mukai92, Kozasa92}. We adopt a fractal dimension $d_f=1.9$ and fractal prefactor $k_0=1.03$, which are the typical values for BCCA aggregates formed with oblique collisions \citep{Tazaki16}. The fractal prefactor is defined by $N=k_0(a_g/a_0)^{d_f}$; $N$ is the number of monomers and $a_g=\sqrt{3/5}a_c$ is the radius of gyration. Since we fix the monomer radius as $a_0=0.1~\mu$m, the free parameter of the model is the aggregate radius $a_c$ only. The volume filling factor is given by $f=3\times10^{-4}(a_c/100~\mu\mathrm{m})^{-1.1}$. In this study, we only consider $a_c>4~\mu$m so that the volume filling factor of fluffy aggregates is always less than 0.01.

\item {\it Compact dust aggregates}: Compact dust aggregates are defined as aggregates with $d_f=3$ and $0.01\le f<1$ in this paper. This model is intended to mimic BPCA aggregates (with/without hierarchical effects).  Aggregate radius and the volume filling factor are the important parameters. We use the term ``compact" because these grains are relatively compact compared to the fluffy dust aggregate model with $f<0.01$. 
\end{itemize}

In this paper, we use the term {\it dust particles}, or simply {\it dust}, in more general contexts, i.e., when we intend to mention more general cases rather than to indicate a specific model given above. 

We mainly focus on dust particle radii from sub-millimeter ($\sim100~\mu$m) to centimeter size  ($\sim10^4~\mu$m). Since this study aims to understand the scattering polarization of disks at (sub)millimeter wavelengths, dust particle radii should be comparable to or larger than the observing wavelength in order to attain high scattering albedo. 

\subsection{Computation of optical properties}
Computing optical properties of compact/fluffy dust aggregates is not an easy task, whereas those of solid spheres can be readily computed by the Mie theory \citep{Bohren83}. 
Powerful numerical techniques for directly solving their optical properties have been developed, such as the T-Matrix Method \citep{Mackowski96} and the Discrete Dipole Approximation \citep[][]{Draine94}.
However, millimeter-sized BPCA aggregates of $0.1~\mu$m monomers contain about $10^{11}$ monomers, and solving for electromagnetic interactions between every pair of monomers is very time-consuming. 
Therefore, computing those properties with numerical techniques is not a realistic choice given current computer capabilities.

In order to reduce computational cost, approximate methods are useful.
For fluffy dust aggregates, we can approximate their optical properties by using the modified mean field theory \citep{Tazaki16, Tazaki18}. On the other hand, the approximation to the optical properties of compact aggregates is somewhat more difficult because multiple scattering comes into play. 
Hence, as our current best, conservative estimate, we adopt the effective medium theory (EMT). 
It was recently found that the optical properties of compact dust aggregates, in particular for scattering opacity, could be (not perfectly but roughly) approximated by EMT \citep[see Figure 6 in][]{Tazaki18}.
This is partly because compact aggregates have $d_f=3$, and hence, their scattering behaviors, such as interference of scattered waves, have some similarities to those of the Mie theory, which also deals with spherical particles ($d_f=3$). Thus, we anticipate that qualitatively similar results can be obtained even if we use EMT. We adopt the Maxwell Garnett mixing rule \citep{MG04} to obtain an effective refractive index, $m_\mathrm{eff}$, of a mixture of vacuum (the matrix component) and dust composition (the inclusion component) described in Section \ref{sec:dustcomp}. Since the Maxwell Garnett mixing rule works well when the matrix component dominates, we only consider $f\le0.1$ \citep{Kolokolova01}.

For solid spheres and compact aggregates, optical properties are averaged over the dust size distribution in order to suppress strong resonances of scattering properties. 
We assume a dust size distribution obeying $n(a)da\propto a^pda$
with maximum and minimum radii of $a_\mathrm{max}$ and $a_\mathrm{min}$, respectively. $n(a)da$ represents the number of dust particles in a size range $[a,a+da]$.
We set $p=-3.5$ and $a_\mathrm{min}=0.01~\mu$m, 
although our results are insensitive to $a_\mathrm{min}$ as long as $a_\mathrm{min}\ll a_\mathrm{max}$.
 
\section{Millimeter-wave scattering properties of dust particles} \label{sec:optprop}
We present optical properties of solid spheres, compact dust aggregates and fluffy dust aggregates at millimeter wavelength. 

It is useful to introduce a quantity to assess an ability of dust particles for producing millimeter-wave scattering polarization.
Dust particles efficiently produce a polarized scattered light when their single scattering albedo $\omega$ is high and the degree of linear polarization $P$ for $90^\circ$ scattered waves is high. Therefore, the product $P\omega$ can be used as a diagnostic of efficient scattering polarization \citep{Kataoka15}.
However, from the point of view of radiative transfer, the single scattering albedo $\omega$ becomes a bad approximation to apparent scattering efficiency 
because forward scattering dominates the value of the albedo when a particle is larger than the wavelength \citep[e.g.,][]{Mulders13, Min16, Tazaki19}. 
Since forward scattering does not seem to change the direction of incident light upon scattering, it is effectively not scattering. In addition, forward scattered light is unpolarized. Thus, it is not so important for polarized radiative transfer. A simple way to mimic apparent scattering efficiency is to define the effective scattering opacity $\kappa_\mathrm{sca}^{\mathrm{eff}}=(1-g)\kappa_\mathrm{sca}$, where $\kappa_\mathrm{sca}$ is the scattering opacity and $g$ is the asymmetry parameter \citep[e.g.,][]{Birnstiel18}. The asymmetry parameter diminishes for isotropic scattering and is unity for perfect forward scattering. 
Thus, if scattered light is highly concentrated into the forward direction $(g\sim1)$, 
$\kappa_\mathrm{sca}^{\mathrm{eff}}\approx0$, indicating that apparently no scattering occurs.
By using the effective scattering opacity, we can also define the effective single scattering albedo $\omega^\mathrm{eff}=\kappa_\mathrm{sca}^{\mathrm{eff}}/(\kappa_\mathrm{sca}^{\mathrm{eff}}+\kappa_\mathrm{abs})$. Therefore, we use $P\omega^\mathrm{eff}$ instead of $P\omega$ in order to quantify millimeter-wave scattering polarization.

\subsection{Solid spheres and Compact dust aggregates} \label{sec:comp}
\begin{figure*}
\begin{center}
\includegraphics[height=6.0cm,keepaspectratio]{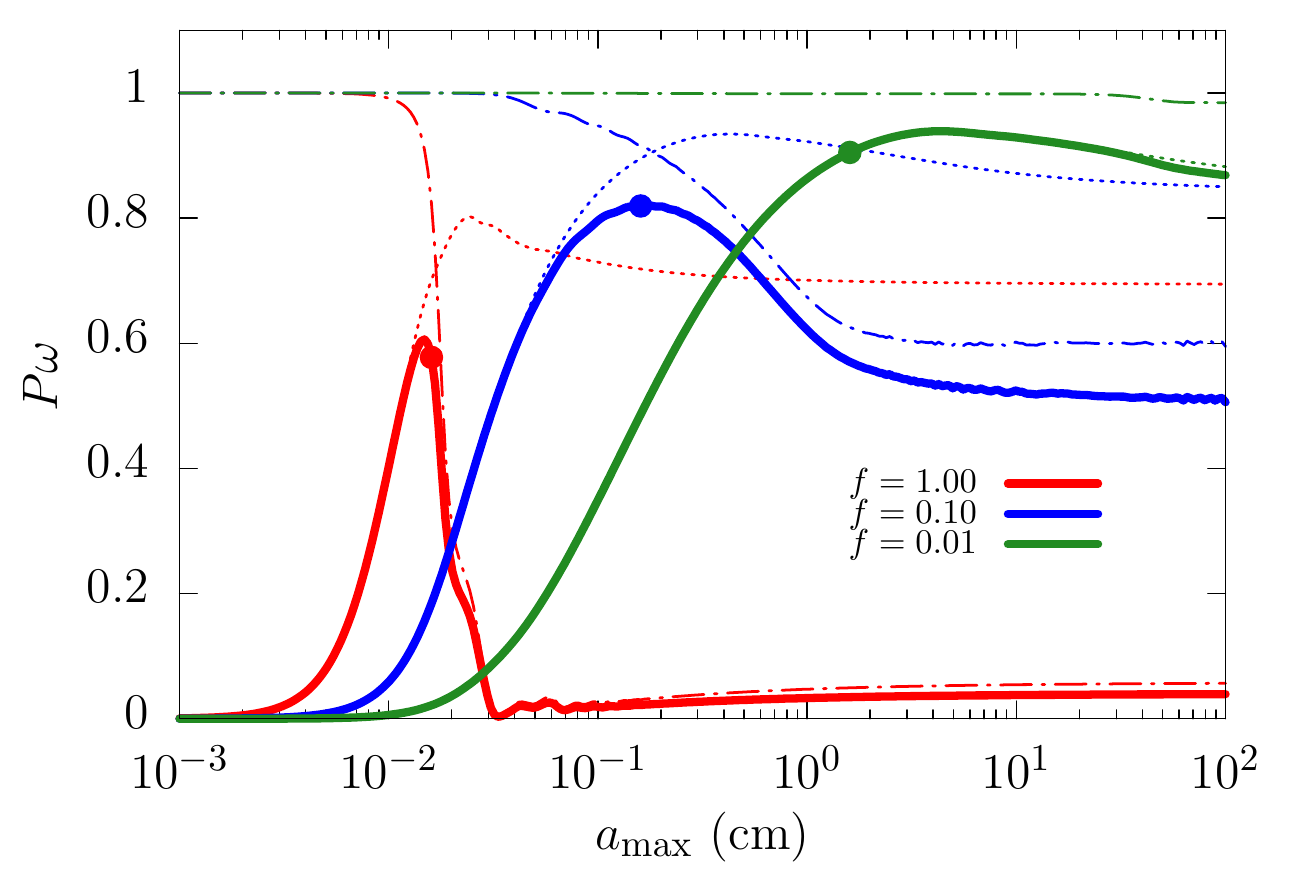}
\includegraphics[height=6.0cm,keepaspectratio]{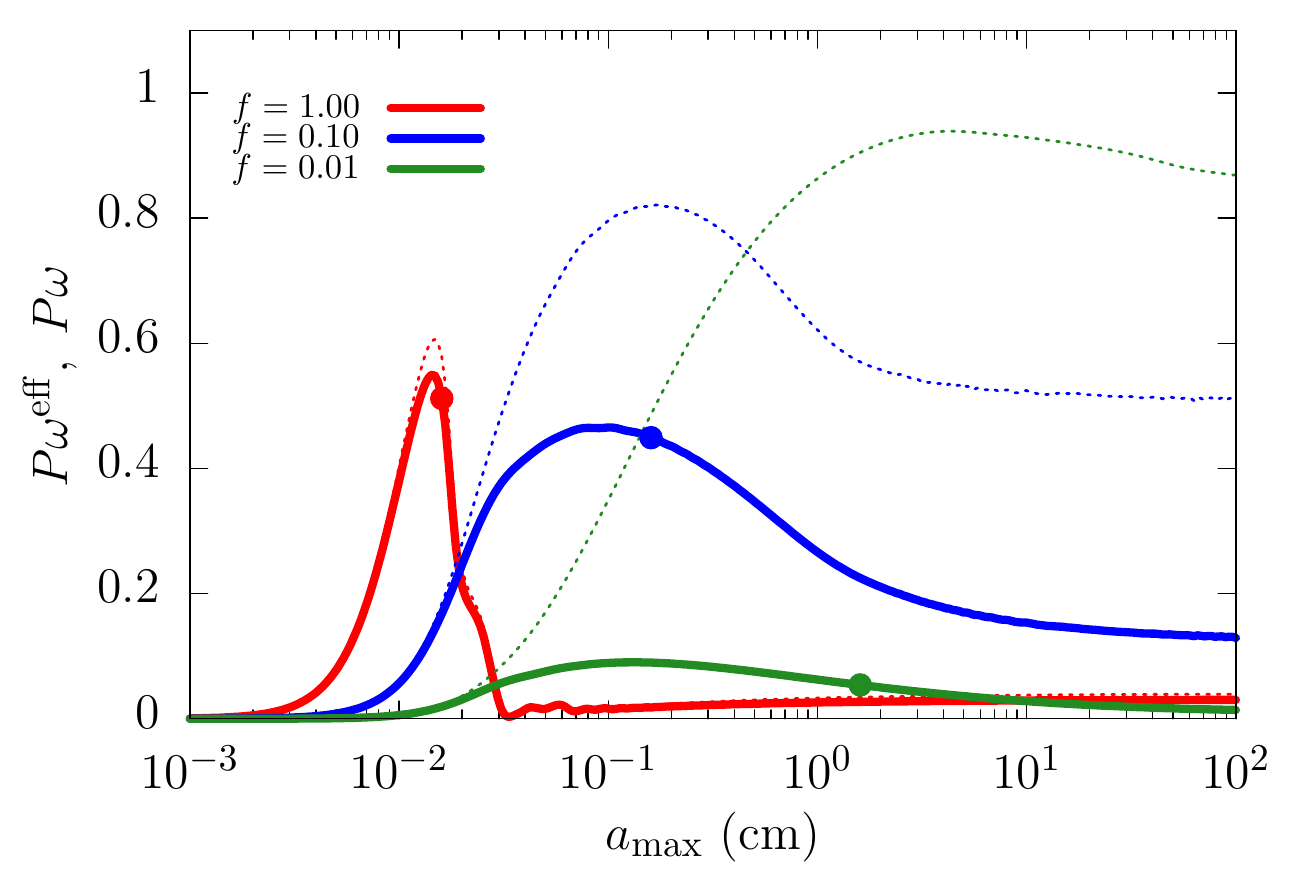}
\caption{$P\omega$ and $P\omega^\mathrm{eff}$ for solid spheres ($f=1$) and compact dust aggregates ($f=0.1,~0.01$) at $\lambda=1$ mm. 
(Left) 
Solid lines represent $P\omega$. Red, blue, and green lines show the results for the volume filling factor of $f=1$, $0.1$, and $0.01$, respectively. Dotted and dot-dashed lines indicate the single scattering albedo $\omega$ and the degree of linear polarization $P$ at scattering angle of 90$^\circ$, respectively. 
Filled circles indicate $P\omega$ at the position of $a_\mathrm{max}f=160~\mu$m.
(Right) Solid and dashed lines indicate results for $P\omega^\mathrm{eff}$ and $P\omega$, respectively. Filled circles indicate $P\omega^\mathrm{eff}$ at $a_\mathrm{max}f=160~\mu$m. The $P\omega^\mathrm{eff}$-values at $x>1$ is significantly attenuated by considering the effective single scattering albedo.}
\label{fig:pomg}
\end{center}
\end{figure*}

Figure \ref{fig:pomg} shows $P\omega$ (left panel) and$P\omega^\mathrm{eff}$ (right panel) as a function of the maximum dust radius $a_\mathrm{max}$ for solid spheres ($f=1$) and compact dust aggregates ($f=0.1,~0.01$) at wavelength $\lambda=1$ mm. 
In order to understand behavior of $P\omega^\mathrm{eff}$ against $a_\mathrm{max}$, we start to discuss $P\omega$, and then $P\omega^\mathrm{eff}$ for each dust model. 
It is also useful to define the size parameter $x=2\pi a_\mathrm{max}/\lambda$. Since we adopt $\lambda=1$ mm, the size parameter of unity corresponds to $a_\mathrm{max}=160~\mu$m.

For solid spheres ($f=1$), the maximum value of $P\omega$ appears at around $x\approx1$. The single scattering albedo $\omega$ is usually much less than unity when $x\ll1$ and becomes of order unity when $x\gtrsim1$. 
On the other hand, the degree of linear polarization $P$ is high as long as $x\ll1$, but it suddenly decreases for $x\gtrsim1$. As a result, $P\omega$ is maximized at $x\approx1$. 
$P\omega^\mathrm{eff}$ for solid spheres is almost the same as $P\omega$ because, as the maximum dust radius increases, the degree of polarization drops much faster than the term $1-g$. Thus, considering $P\omega^\mathrm{eff}$ instead of $P\omega$ does not significantly change the previous interpretation by \citet{Kataoka15}, where solid spherical particles are assumed.

Next, we focus on compact dust aggregates ($f=0.1$ and $0.01$). 
First of all, decreasing the volume filling factor leads to a reduction of the single scattering albedo at $x\le1$ because the scattering opacity is proportional to the aggregate mass, whereas the absorption opacity is not. 
A notable impact of decreasing $f$ is that the degree of polarization becomes high even if $x>1$. 
This is due to the fact that aggregates with $f\ll1$ have an effective refractive index close to that of vacuum, and are regarded as optically thin particles. Since multiple scattering is suppressed inside optically thin particles, a high degree of polarization is obtained. $P\omega$ increases with increasing $a_\mathrm{max}$ until $x|m_\mathrm{eff}-1|$ exceeds unity at which aggregates become optically thick. Since $x|m_\mathrm{eff}-1|$ is proportional to $a_\mathrm{max}f$ \citep{Kataoka14}, the maximum value of $P\omega$ approximately occurs at $a_\mathrm{max}f\approx\lambda/2\pi\approx~160~\mu$m. 
For example, when $f=0.1$, $P\omega$ is maximized at approximately $a_\mathrm{max}=1.6$ mm. 
The maximum $P\omega$-value is larger for a lower filling factor because lower filling factor makes aggregates less absorbing and also suppresses depolarization due to multiple scattering.

Unlike the case of solid spheres, the effective single scattering albedo plays a significant role for compact aggregates. As shown in Figure \ref{fig:pomg} (right), the maximum value of $P\omega^\mathrm{eff}$ is significantly attenuated at $x>1$, in particular for the case of $f=0.01$.

As a result, as the volume filling factor decreases, the maximum value of $P\omega^\mathrm{eff}$ decreases. This suggests that compact aggregates with lower filling factor produce fainter polarized-scattered light than solid spheres. We will confirm this by performing radiative transfer simulations in Section \ref{sec:result}.

\subsection{Fluffy dust aggregates} \label{sec:fracl}
The scattering properties of fluffy dust aggregates are significantly different from those of spherical particles. We use the MMF theory to compute optical properties of fluffy aggregates. In this section, we use $d_f=2.0$ and $k_0=1.0$ in order to attain clear physical insights into the optical properties of fluffy dust aggregates.

First of all, the degree of linear polarization of fluffy dust aggregates is as high as $100\%$ because multiple scattering is suppressed when $x_0\ll1$ and $d_f\le2$ \citep{Tazaki16}, where $x_0=2\pi a_0/\lambda$ is the size parameter of the monomer particle.
Here, the wavelength of interest is (sub-)millimeter wavelength and the monomer radius is of sub-micron size, and therefore, this condition is satisfied. Even if $a_c\gg\lambda/2\pi$, the highest degree of polarization remains $100\%$. Thus, we obtain $P\omega^\mathrm{eff}=\omega^\mathrm{eff}$.

\begin{figure}
\begin{center}
\includegraphics[height=6.0cm,keepaspectratio]{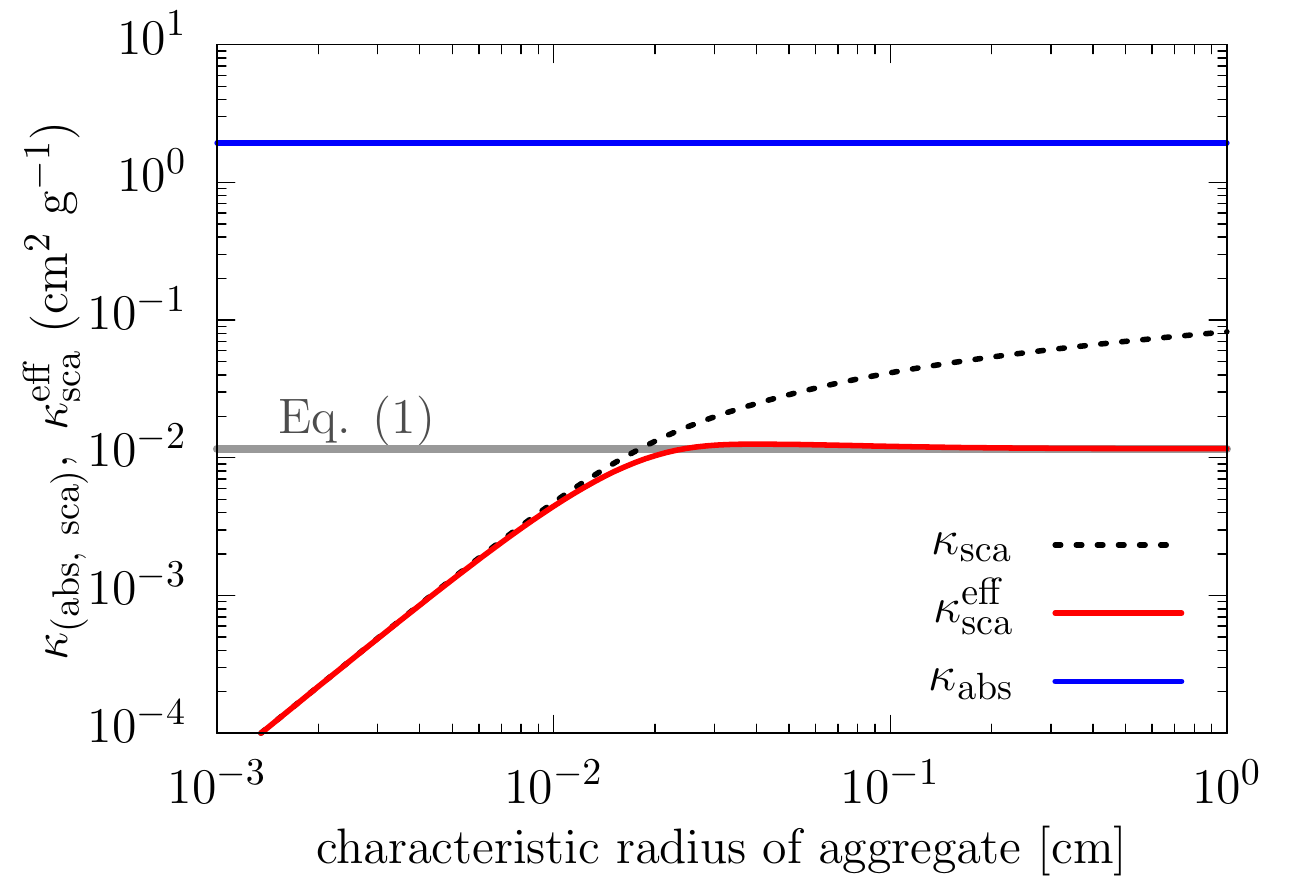}
\caption{Opacities of fluffy dust aggregates with fractal dimension $d_f=2$ and fractal prefactor $k_0=1.0$ at $\lambda=1$ mm. Red and blue lines represent the effective scattering opacity and the absorption opacity, respectively. The dashed line indicates the scattering opacity. The gray horizontal line is an analytical upper limit (Equation \ref{eq:albd}).}
\label{fig:mmf_opac}
\end{center}
\end{figure}

In Figure \ref{fig:mmf_opac}, we show the opacities of fluffy dust aggregates as a function of the characteristic radius $a_c$. The absorption opacity does not depend on $a_c$ because fluffy aggregates with different $a_c$ have the same mass-to-area ratio, and the absorption opacity remains the same \citep{Kataoka14} \footnote{Strictly speaking, in the Rayleigh limit, the absorption opacity depends on the number of monomers \citep{Stognienko95, Henning96, Tazaki18} due to monomer-monomer interaction. However, this effect is saturated when the number of monomers exceeds $\sim100$, and then the size dependence becomes negligible for further large aggregates.}. 
At $a_c\lesssim\lambda/2\pi$, the scattering opacity increases with increasing aggregate radius (mass). On the other hand, once the aggregate radius exceeds $\lambda/2\pi$, the effective scattering opacity saturates. An analytical solution to the upper limit on the effective scattering opacity can be found under the conditions of $x_0\ll1$ and $d_f=2$ (see Appendix \ref{sec:appB} for derivation). 
The analytical solution to the effective scattering opacity of fluffy dust aggregates at $a_c>\lambda/2\pi$ is
\begin{eqnarray}
\kappa_\mathrm{sca}^\mathrm{eff}&=&\frac{\kappa_{\mathrm{sca,mono}}}{2x_0^2},\label{eq:albd}
\end{eqnarray}
where $\kappa_{\mathrm{sca,mono}}$ is the scattering opacity of the single monomer.
It is clear from Equation (\ref{eq:albd}) that the effective scattering opacity of fluffy aggregates with $d_f=2$ does not depend on the aggregate radius or the number of monomers. 

\begin{figure}
\begin{center}
\includegraphics[height=6.0cm,keepaspectratio]{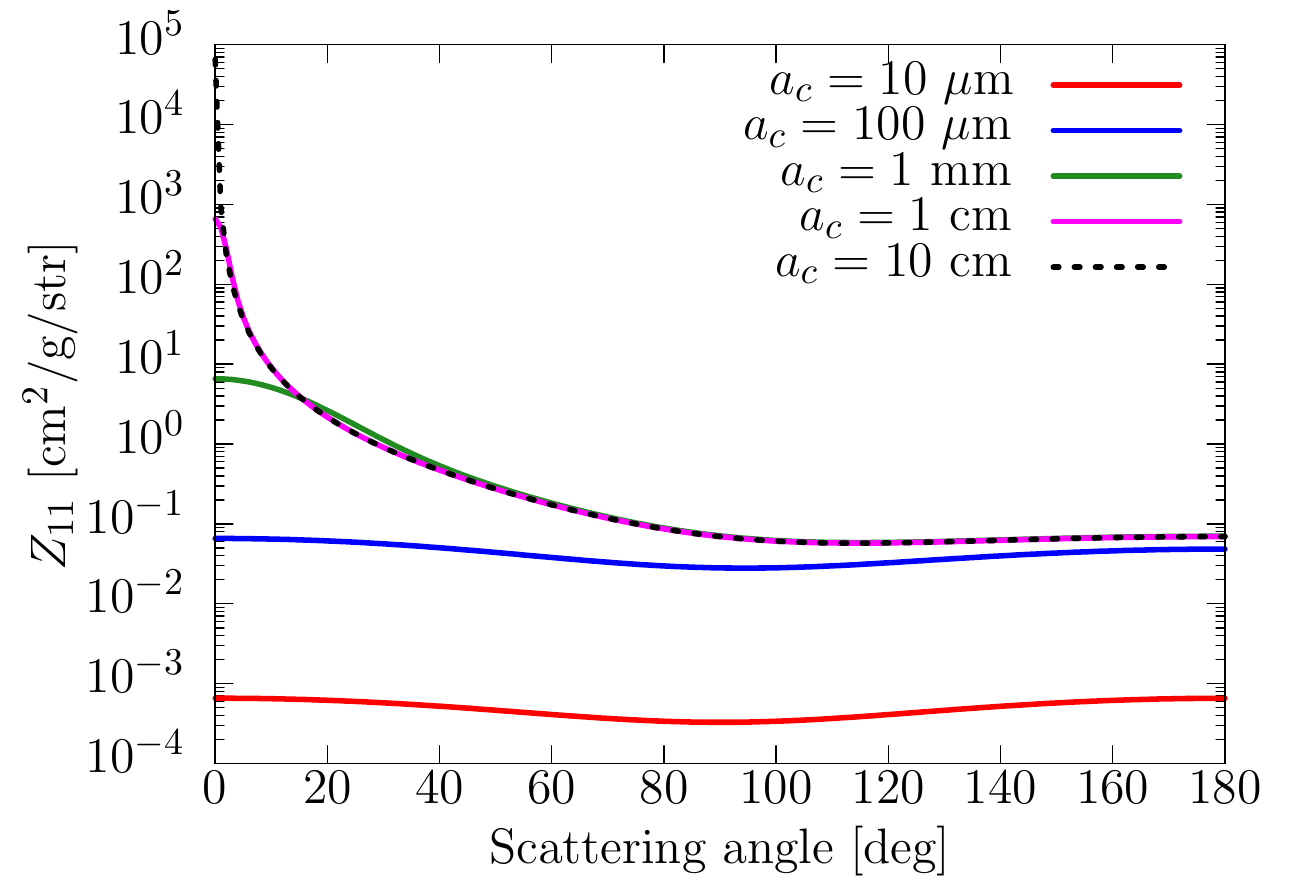}
\caption{Differential scattering cross-section per unit mass of fluffy dust aggregates with $d_f=2$ and $k_0=1.0$ at $\lambda=1$ mm. As the aggregate radius increases, the differential cross-section saturates at intermediate and backward scattering angles as a natural consequence of single scattering.}
\label{fig:mmf_z11}
\end{center}
\end{figure}

The physical interpretation of Equation (\ref{eq:albd}) can be captured when the differential scattering cross-section per unit mass $Z_{11}$ is plotted as a function of scattering angle (Figure \ref{fig:mmf_z11}). For an aggregate smaller than the wavelength, Rayleigh scattering occurs. This means that scattering is coherent for all scattering angles.
Once the aggregate radius becomes larger than the wavelength, $Z_{11}$ saturates at intermediate and backward scattering angles. 
This is because scattered waves from a pair of monomers separated by a distance larger than (approximately) the wavelength are out-of-phase. On the other hand, scattered waves from particles separated by smaller distance can be in phase for such scattering angles.
Therefore, scattered light at these scattering angles is dominated by coherent scattered light from the small-scale structure of the aggregate rather than the large-scale structure. 
As a consequence, $Z_{11}$ becomes irrelevant to $a_c$ at these scattering angles, and then $\kappa_\mathrm{sca}^\mathrm{eff}$ also becomes independent on how large the aggregate is.
The above explanation is also illustrated schematically in Figure 7 in \citet{Tazaki19}. 

By using Equation (\ref{eq:albd}) and formulae for opacities in the Rayleigh limit \citep{Bohren83}, the effective single scattering albedo of fluffy aggregates, whose characteristic radius is larger than the wavelength, is
\begin{equation}
\omega^\mathrm{eff}=\frac{x_0}{x_0+3\mathcal{R}},
~\mathcal{R}=\mathrm{Im}\left\{\frac{m^2-1}{m^2+2}\right\}\bigg/\left|\frac{m^2-1}{m^2+2}\right|^2. \label{eq:oeffana}
\end{equation}
By substituting values into Equation (\ref{eq:oeffana}), we obtain $P\omega^\mathrm{eff}=6\times10^{-3}$, where we have assumed $m=2.6+0.074i$ and $P=1$. As a result, it is found that fluffy aggregates are inefficient scatterers of millimeter-wave radiation. Even if the monomer radius is a few microns, the effective albedo is still $\lesssim0.1$. Therefore, fluffy aggregates of (sub-)micron-sized monomers are not likely to contribute millimeter-wave scattering in protoplanetary disks. 


\begin{figure}
\begin{center}
\includegraphics[height=6.0cm,keepaspectratio]{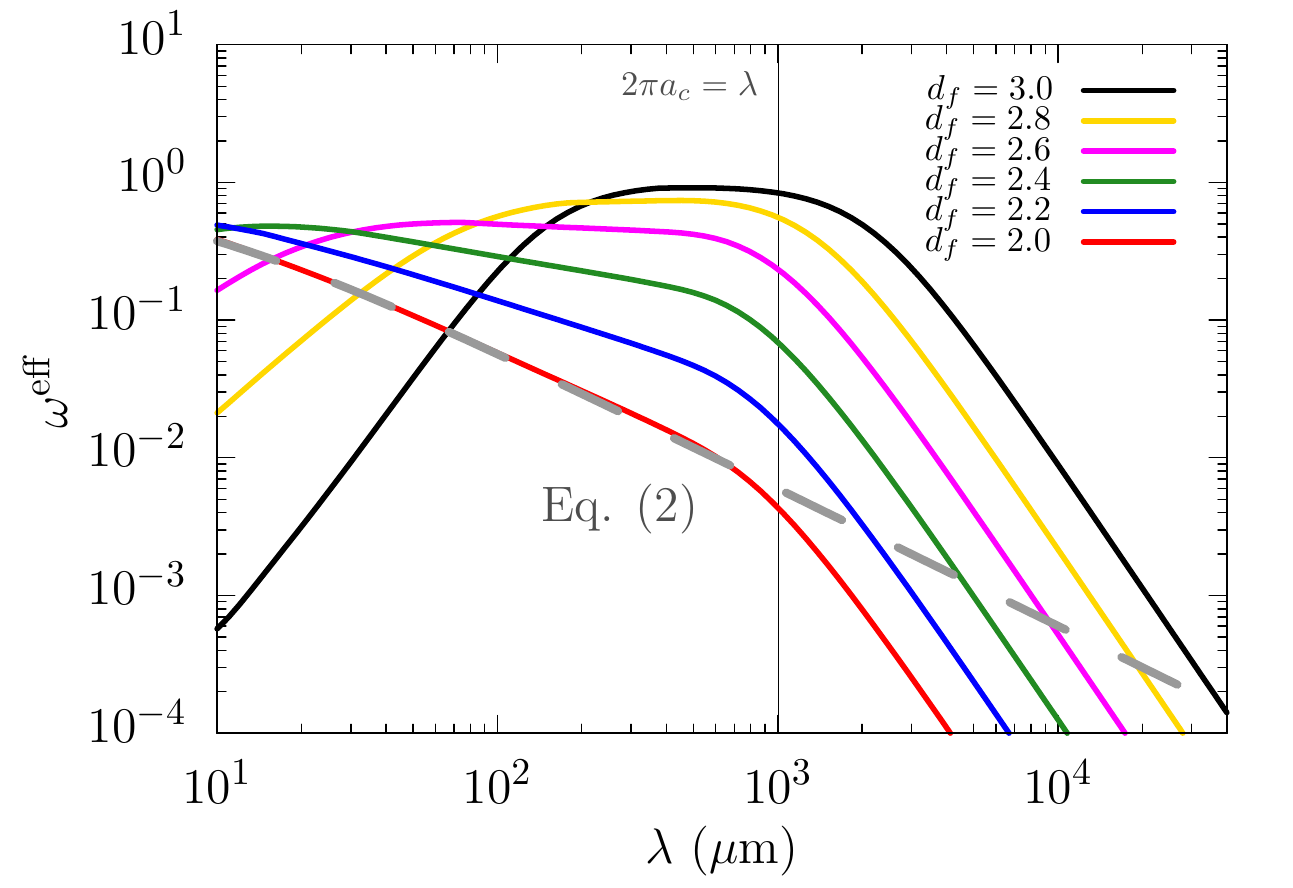}
\caption{The effective single scattering albedo $\omega^\mathrm{eff}$ of dust aggregates with $a_c=160~\mu$m for various $d_f$ estimated by the MMF theory. Dashed line represent analytic solution applicable when $\lambda<2\pi{a_c}$ and $d_f=2$ (Equation \ref{eq:oeffana}). The vertical line indicates the position of $\lambda=2\pi a_c$. Refractive index is set as $m=2.6+0.074i$.}
\label{fig:intdf}
\end{center}
\end{figure}
Finally, we address how a different choice of fractal dimension affects scattering properties, since the fractal dimension of dust aggregates in disks might be $2\lesssim d_f \lesssim 3$. 
Although the detailed angular dependence of scattering matrix elements is not easy to compute, the qualitative wavelength dependence of opacities might be obtained by using the MMF theory \citep{Tazaki18}. In Figure \ref{fig:intdf}, we show the effective scattering albedo of dust aggregates with various fractal dimension having $a_c=160~\mu$m and $a_0=0.1~\mu$m. Since the MMF theory is incapable of predicting the degree of polarization $P$ for $d_f>2$ due to the importance of multiple scattering, we only show $\omega^\mathrm{eff}$ in Figure \ref{fig:intdf}.  

At $\lambda>2\pi a_c=1$ mm (Rayleigh limit), the effective albedo increases with fractal dimension. Since we have fixed both $a_c$ and $a_0$, increasing fractal dimension results in increasing the aggregate mass. Therefore, the effective albedo increases with $d_f$.
At $\lambda\ll2\pi a_c$, the effective albedo for higher fractal dimension becomes smaller due to forward scattering. Figure \ref{fig:intdf} predicts that aggregates with higher fractal dimension produce faint and reddish scattered light in the short wavelength domain, e.g., infrared wavelengths, whereas those with $d_f=2$ produce bright and blue scattered light. 
This tendency has already been confirmed by rigorous computations of optical properties as well as radiative transfer simulations \citep{Mulders13, Min16, Tazaki19}.

As a result, at (sub-)millimeter wavelengths, increasing the fractal dimension of aggregates gradually increases the effective scattering albedo. Thus, aggregates with higher fractal dimension are more likely to scatter at millimeter wavelength. 

\section{Radiative Transfer Simulations} \label{sec:result}
In Section \ref{sec:optprop} we discussed the optical properties of solid spheres and compact/fluffy dust aggregates. In this section we perform radiative transfer simulations of disks containing these dust particles.

\begin{table*}
\caption{Optical Properties of Dust models at Wavelength $\lambda=1$ mm}
\label{tab:r}
\centering
\begin{tabular}{llcllcllcllcllcll}
\hline
${\rm Dust\ model}$ && $\kappa_\mathrm{abs}\ (\mathrm{cm}^2\ \mathrm{g}^{-1})$ & $\kappa_\mathrm{sca}^\mathrm{eff}\ (\mathrm{cm}^2\ \mathrm{g}^{-1})$& $g$ & $\omega^\mathrm{eff}$ & $P(\theta={90}^\circ)$  & $P\omega^\mathrm{eff}$\\
\hline  \hline
Solid Spheres ($a_\mathrm{max}=160\ \mu\mathrm{m}$) && 3.37 & 5.54 &  0.30 & 0.62 & 0.82 & 0.51 \\
Compact aggregates ($a_\mathrm{max}=1.6\ \mathrm{mm}$, $f=0.1$) && 2.17 & 2.09&  0.89 & 0.49 & 0.92 & 0.45 \\
Compact aggregates ($a_\mathrm{max}=1.6\ \mathrm{cm}$, $f=0.01$) && 1.95 & 0.11&  0.99 & 0.05 & 1.00 & 0.05 \\
Fluffy aggregate ($a_c=160\ \mu\mathrm{m}$) && 1.94 & $4.29\times10^{-3}$ &  0.14 & $2.21\times10^{-3}$ & 1.00 & $2.21\times10^{-3}$ \\
\hline
\hline
\end{tabular}
\label{tab:optprop}
\end{table*}

\subsection{Model and Method}
In order to simulate millimeter-wave scattering polarization of a protoplanetary disk, we use a publicly available 3D Monte Carlo radiative transfer code RADMC-3D \citep{Dullemond12}.
We assume a vertically isothermal disk, and the radial temperature profile approximately obeys $T(r)=86\ \mathrm{K}(r/10\ \mathrm{au})^{-0.5}$, where $T$ is the dust temperature and $r$ is the distance from the central star. The inner and outer disk radii are 10 au and 100 au, respectively, and the dust surface density is set as $\Sigma_\mathrm{d}=0.157~\mathrm{g}~\mathrm{cm}^{-2}~(r/10\ \mathrm{au})^{-1}$, leading to $10^{-4}M_\odot$ total disk dust mass. We assume the dust model is the same everywhere within the disk. In addition, we ignore polarized thermal emission from aligned grains. 
The number of photon packets used in Monte Carlo scattering simulations is $10^9$.
Since our primary focus is on how dust properties affect polarization, we fix both the disk model and disk inclination angle. 
Since moderate disk inclination angles are favored for scattering polarization \citep{Yang16a}, we assume the inclination angle is 45$^\circ$. The imaging wavelength is set as $\lambda=1~$mm.

Dust models used in simulations are solid spheres and compact/fluffy dust aggregates.
In Table \ref{tab:optprop}, we summarize the optical properties used in simulations for solid spheres and compact dust aggregates ($a_\mathrm{max}f=160~\mu$m with $f=1,0.1,0.01$) as well as for fluffy dust aggregates ($d_f=1.9$, $k_0=1.03$, $a_c=160~\mu$m). It is worth reminding the reader that the value of $160~\mu$m is chosen so that the dust particle radius is equal to $\lambda/2\pi$, where $\lambda=1~$mm. We will also discuss how different choices of $a_\mathrm{max}$ affect the results.

The quantity $a_\mathrm{max}f$ is a useful quantity to characterize particle properties because the mass-to-area ratio of dust particles is proportional to $af$, and thus, both absorption and dynamical properties are characterized by $af$ \citep{Kataoka14}. Indeed, as pointed out by \citet{Kataoka14}, the mass absorption opacity of aggregates with the same $a_\mathrm{max}f$-value become very similar (Table \ref{tab:optprop}).

\subsection{Results of radiative transfer simulations} \label{sec:respo}
\begin{figure*}
\begin{center}
\includegraphics[height=6.5cm,keepaspectratio]{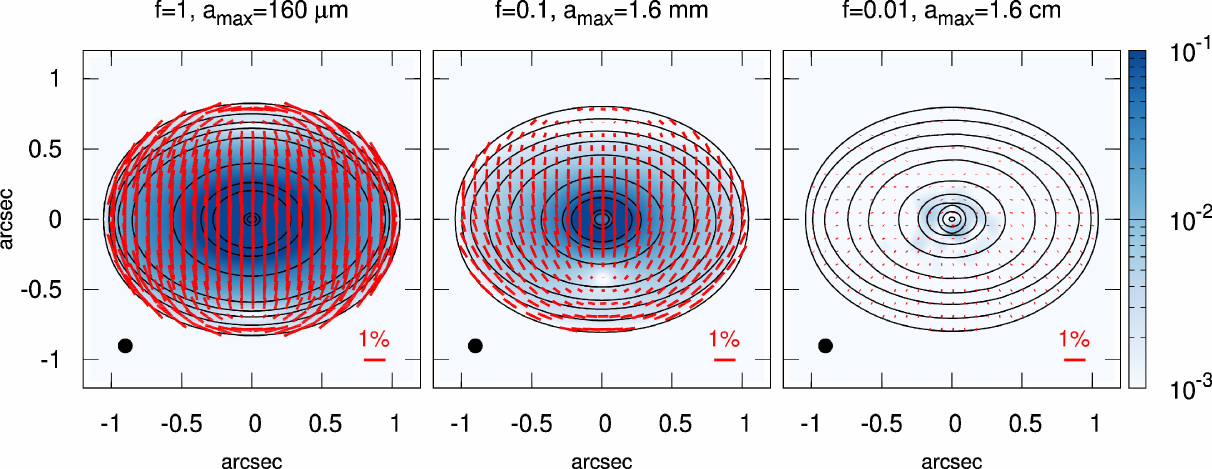}
\caption{
Polarized intensity [mJy/beam] at $\lambda=1$ mm for solid spheres ($f=1$) and compact dust aggregates ($f=0.1$, 0.01) with $a_\mathrm{max}f=160\ \mu$m. From left to right panel, the volume filling factor is decreased. Red bars and their length represent polarization orientations and polarization fraction, where the reference length for $1$\% polarization fraction is shown in right bottom in each panel.
Total intensity contours in each panel are shown for $[3,12,25,35,50,100,200,300] \times 0.03$ [mJy/beam]. FWHM of the beam size is 0".1 as shown in the left bottom circle. Inclination angle of the disk is 45 degrees and the distance between the observer and the disk is assumed to be 100 pc. Bottom side of each image corresponds to the near side of the disk.
}
\label{fig:result_emt}
\end{center}
\end{figure*}

Figure \ref{fig:result_emt} shows a polarization map of a disk containing solid spheres ($f=1$) and compact aggregates ($f=0.1,~0.01$). In each image, polarization angles (E-vector direction) are indicated by bars whose length is proportional to the polarization fraction.

As the volume filling factor decreases, the polarized intensity from the disk diminishes. This can be interpreted as a consequence of low effective scattering albedo as shown in Figure \ref{fig:pomg}.
For the cases of $f=1$ and $0.1$, polarization angles (E-vector direction) tend to be oriented parallel to the disk minor axis at the inner regions of disks. This is characteristic of scattering polarization from an optically thin disk \citep{Kataoka16, Yang16a}.

For $f=0.1$, near-and-far side asymmetry of the polarization pattern can be seen, whereas, for $f=1$, the pattern is symmetric. 
For $f=0.1$, the polarization orientation at the disk near side tends to be orientated in the azimuthal direction, whereas that of the far-side is orientated parallel to the minor axis. The dip in polarized intensity at the near side is the location where the polarization angle changes its direction from radial at the inner disk to azimuthal at the outer disk. 
This asymmetric polarization pattern is caused by anisotropic scattering.
For $f=1$, the assumed particle radius is comparable to $\lambda/2\pi$, and hence, scattering is close to isotropic. However, for $f=0.1$, the assumed particle radius ($a_\mathrm{max}=1.6$ mm for the $f=0.1$ model) is larger than $\lambda/2\pi\approx 160~\mu$m, and thus, forward scattering occurs. Since forward scattered light amplifies a scattered-light component with azimuthal polarization at the disk near-side, near-far side asymmetry of the polarization pattern appears. 

\begin{figure}
\begin{center}
\includegraphics[height=6.0cm,keepaspectratio]{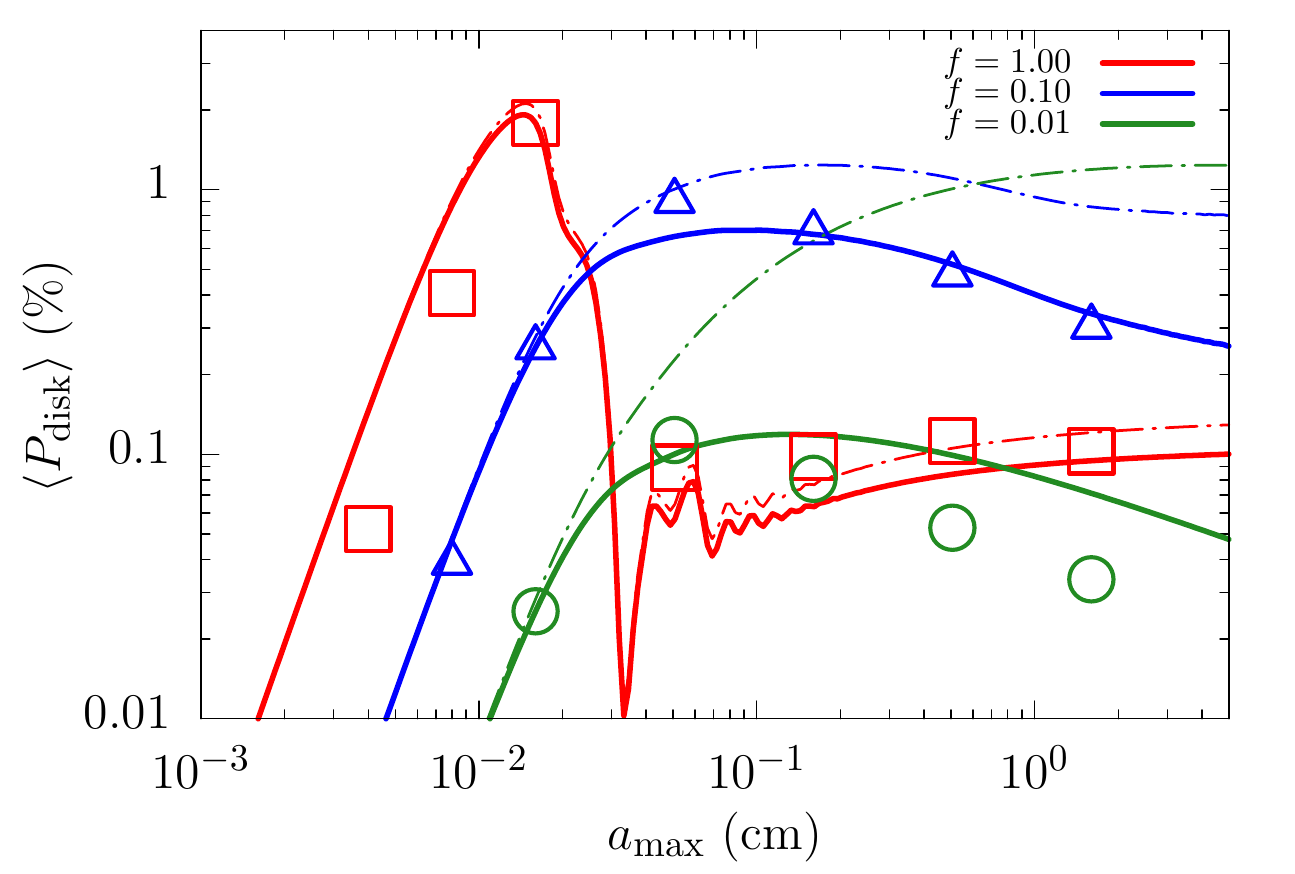}
\caption{
Disk polarization fraction $\langle P_\mathrm{disk} \rangle$ versus $a_\mathrm{max}$ for $f=1$ (red), $0.1$ (blue), and $0.01$ (green). 
The square, triangle, and circle symbols show the values of $\langle P_\mathrm{disk} \rangle$ directly obtained from radiative transfer simulations, while the solid and dot-dashed lines show the estimates $\langle P_\mathrm{disk} \rangle=CP\omega^{\mathrm{eff}}$ and $\langle P_\mathrm{disk} \rangle=CP\omega$, respectively.
}
\label{fig:pomgeff_comp}
\end{center}
\end{figure}

Next, we study the dust particle radius dependence of the average polarization fraction obtained by radiative transfer simulations.
Figure \ref{fig:pomgeff_comp} shows the average polarization fraction $\langle P_\mathrm{disk} \rangle$, which is a ratio of the polarized flux to the total flux for the entire disk, for various $a_\mathrm{max}$.
We compare the particle radius dependence of $P\omega^{\mathrm{eff}}$ as well as $P\omega$, where we use $\langle P_\mathrm{disk} \rangle=CP\omega^{\mathrm{eff}}$ to compare simulation results; the $C$-value is a numerical factor calibrated at $a_\mathrm{max}=160\ \mu$m for each model.
 
As shown in Figure \ref{fig:pomgeff_comp}, the solid sphere model ($f=1$) shows strong dependence on $a_\mathrm{max}$. This means that a high polarization fraction occurs at the vicinity of  $a_\mathrm{max}\approx\lambda/2\pi$. 
On the other hand, for the case of $f=0.1$, dependence on dust radius becomes much weaker. Therefore, when dust porosity is taken into account, scattering polarization can be detected for a wider range of dust particle radius compared to the solid sphere model, although if the porosity is too high ($f=0.01$) insufficient polarized intensity is produced.
At $a_\mathrm{max}=160~\mu$m, decreasing the volume filling factor makes the polarization fraction of scattered light small as a consequence of lower albedo for lower filling factor particles. This is consistent with the results in Figure \ref{fig:pomg}.

As shown in Figure \ref{fig:pomgeff_comp}, $P\omega^\mathrm{eff}$ can reproduce the overall dependence of polarization fraction on $a_\mathrm{max}$ observed in simulations.
It is also shown that $P\omega$ fails to reproduce the dust particle radius dependence, indicating the importance of considering $\omega^\mathrm{eff}$ instead of $\omega$.

As a result, we conclude that dust aggregates with lower filling factor produce smaller polarization fractions due to the effective reduction of their scattering opacity caused by strong forward scattering. In other words, higher polarization fraction is more likely to be produced by relatively compact dust particles ($f\gtrsim0.1$).

\begin{figure}
\begin{center}
\includegraphics[height=6.0cm,keepaspectratio]{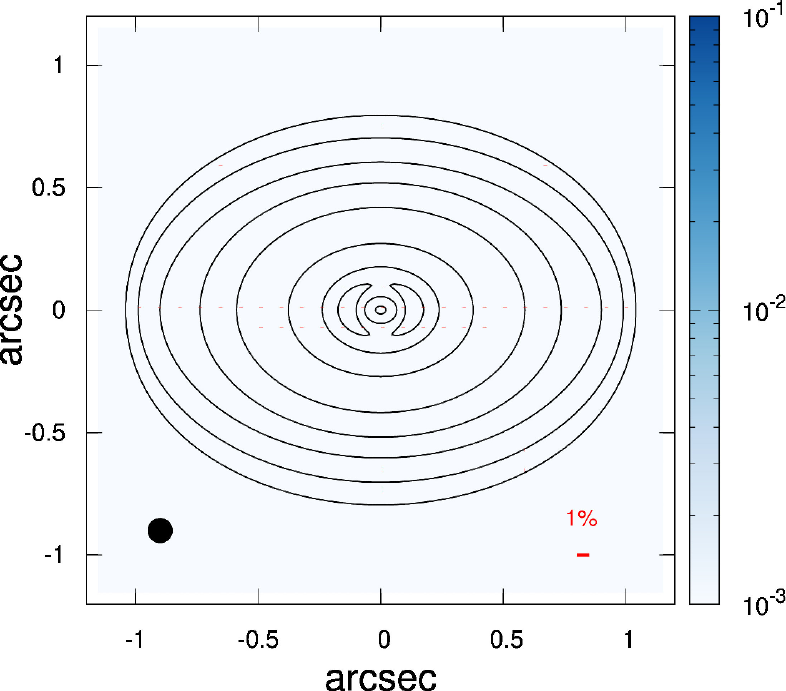}
\caption{Same as Figure \ref{fig:result_emt}, but for fluffy dust aggregates with $a_c=160\ \mu$m. Polarized intensity is much fainter than that of solid sphere and compact aggregate models.}
\label{fig:bcca}
\end{center}
\end{figure}

Next, we perform radiative transfer simulations with fluffy dust aggregates with $a_c=160~\mu$m, and the result is shown in Figure \ref{fig:bcca}. It is found that fluffy aggregates show very faint polarized intensity. This is mainly due to low scattering albedo as can be seen in Table \ref{tab:r}. What happens when the aggregate radius is further increased? We perform additional radiative transfer simulations for fluffy aggregates with $a_c=1.6$ mm and $1.6$ cm, and we find that these larger aggregates give rise to almost the same results as found with $a_c=160~\mu$m. This is due to saturation of the effective scattering opacity as can be seen in Figure \ref{fig:mmf_opac}. As a result, it is found that a fluffy aggregate of sub-micron monomers with any characteristic radius is unlikely to produce millimeter-wave scattering polarization.

As a result, we conclude that higher fractal dimension and higher filling factor are more favorable to producing millimeter-wave scattering polarization.

\section{Discussion} \label{sec:disc}
By comparing our results (Sections \ref{sec:optprop} and \ref{sec:result}) and ALMA polarimetric observations, Section \ref{sec:alma} discusses constraints on dust size and porosity in protoplanetary disks. We also discuss the wavelength dependence of scattering polarization and implications for future multi-wavelength polarimetric observations. In Section \ref{sec:impl}, we discuss implications for planetesimal formation.
In Section \ref{sec:67p}, we also compare our results with the cometary dust particles of 67P/Churyumov-Gerasimenko.

\subsection{Comparison with disk observations} \label{sec:alma}
%
%
\subsubsection{Dust structure and porosity estimate} \label{sec:constf}
Recent polarimetric observations of disks by ALMA have shown that scattering polarization can be commonly seen in various disks. In particular, observed scattering-like polarization (polarization angles parallel to the disk minor axis) often shows a polarization fraction of about 1\% \citep{Kataoka16b, Stephens17, Hull18, Lee18, Bacciotti18, Girart18, Ohashi18, Dent19, Harrison19}. 

In Section \ref{sec:result}, we found that 
compact aggregates with $f=0.01$ and fluffy aggregates ($f\ll0.01$) do not produce polarized-scattered waves at millimeter wavelength. 
In other words, aggregates with extremely high porosity ($f\lesssim0.01$ or porosity higher than 99\%) 
do not explain observations whatever the value of their fractal dimension is.
Dust particles with higher fractal dimension ($d_f\gtrsim2$) and/or higher volume filling factor ($f\gtrsim0.1$) seem to be necessary to explain the observed scattering polarization. 
In order to constrain $f$ and $d_f$ in more detail, further radiative transfer modeling for each object is necessary, although this is beyond the scope of this paper.

It is important to keep in mind that we cannot rule out the presence of dust particles with lower fractal dimension ($d_f\lesssim2$) and/or lower volume filling factor ($f\lesssim0.1$) in disks because these particles are just invisible in millimeter-wave scattering.
Mixed populations of fluffy aggregates and compact aggregates might be another solution to explain observations. However, in this case, increasing the mass abundance of fluffy aggregates with respect to compact aggregates will reduce the polarization fraction because fluffy aggregates only contribute to the total flux via thermal emission. 
Hence, a large mass abundance of fluffy dust aggregates might not be favored.

We present a rule-of-thumb estimate of the upper limit of the mass abundance of fluffy dust aggregates. Suppose a disk consists of two dust populations: solid spheres and fluffy dust aggregates, and denote $M_s$ and $M_f$ by the total mass of solid spheres and fluffy dust aggregates in the disk. If their absorption opacities are similar and the disk is optically thin, the polarized intensity is proportional to $M_s$, while the total intensity is proportional to $M_f+M_s$. The polarization fraction of the disk, $P_\mathrm{disk}$, may be approximated by $P_\mathrm{disk}\approx P_0 M_s/(M_s+M_f)$, where $P_0$ is the polarization fraction of the disk consisting of solid spheres only. In our simulation, $P_0\simeq2.4$ \% (Figure \ref{fig:result_emt}). Since observed polarization fraction is about 1\%, the mass abundance of fluffy aggregates should be $M_f/(M_f+M_s)\lesssim0.6$. Although this upper limit depends on a disk model used, such as temperature structure and optical depth, more detailed analysis is necessary. However, this is beyond the scope of this paper. 
In any case, we at least need a population of dust particles with relatively compact structure to explain polarized-scattered waves. 

The presence of fluffy aggregates could be tested by investigating polarized thermal emission from aligned grains in disks. In disks, in addition to the scattering polarization studied in this paper, polarized thermal emission due to grain alignment has also been proposed  \citep{Cho07, Tazaki17, Bertrang17, Yang19, Kataoka19}. Recently, \citet{Kirsch19} investigated the intrinsic polarization properties of porous dust particles.
They found that a porosity higher than 70\% is not favorable to produce polarized thermal emission. Their conclusion is similar to those obtained in this study.

\subsubsection{Dust size estimate}
Hereafter, we assume that dust particles in disks have $d_f=3$, since higher fractal dimension is favored.
For dust particles with $d_f=3$, the polarization fraction depends on dust radius as we showed in Figure \ref{fig:pomgeff_comp}. Conversely, the observed scattering polarization fraction should contain information from which we can constrain the dust radius in disks.

Since previous studies have relied on the assumption of $f=1$ (solid sphere), scattering polarization occurs efficiently at the vicinity of $a_\mathrm{max}\approx\lambda/2\pi$ (Figure \ref{fig:pomgeff_comp}). 
Thus, the derived maximum dust radius is about a few times $100~\mu$m because the observing wavelengths of ALMA are in the (sub-)millimeter domain.
However, the maximum dust radius is thought to depend on the physical properties of disks, such as age, gas and dust surface density \citep{Birnstiel12}. Hence, if the derived dust constraints are true, we need to explain how $a_\mathrm{max}$ of various disks is fine-tuned at about this radius.

\begin{figure}
\begin{center}
\includegraphics[height=6.0cm,keepaspectratio]{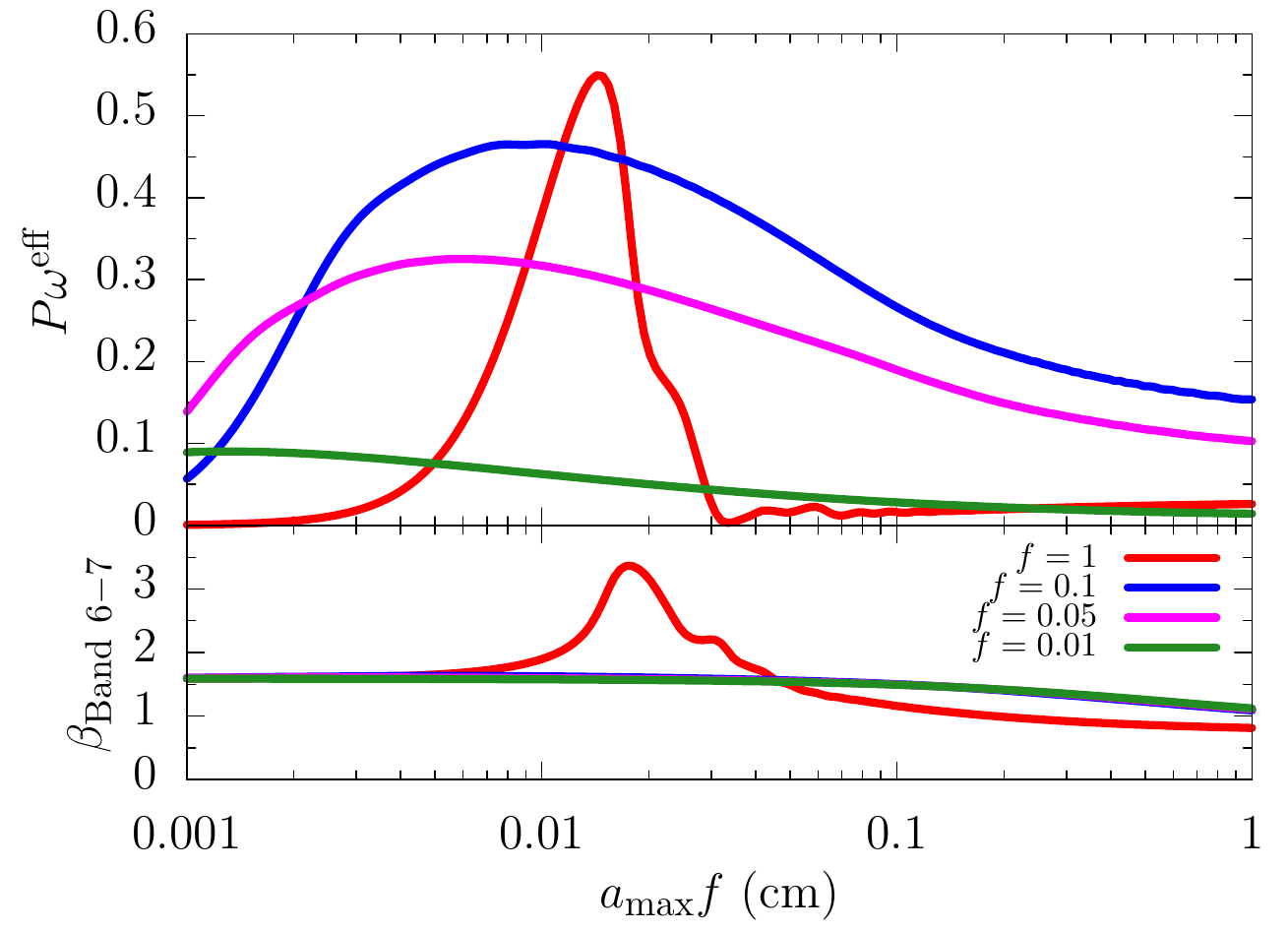}
\caption{Comparison of scattering polarization efficiency $P\omega^\mathrm{eff}$ (top) and opacity index $\beta$ (bottom) as a function of $a_\mathrm{max}f$. Wavelength is set as 1 mm. Different lines represent different volume filling factor $f=1, 0.1, 0.05,$ and $0.01$.}
\label{fig:beta}
\end{center}
\end{figure}

By considering particle porosity, this fine-tuning issue may be relaxed. 
Figure \ref{fig:beta} shows $P\omega^\mathrm{eff}$ as a function of $a_\mathrm{max}f$ for various values of the volume filling factor. For the case of $f=1$, $P\omega^\mathrm{eff}$ is a sharp function of $a_\mathrm{max}$, where the peak appears at $a_\mathrm{max}=\lambda/2\pi$. 
As the volume filling factor decreases, the width of $P\omega^\mathrm{eff}$ broadens, while the peak value is attenuated (see also Figure \ref{fig:pomgeff_comp}).
Therefore, for particles with moderate porosity ($0.1\lesssim f \lesssim 1$), scattering polarization can be expected for a wider range of $a_\mathrm{max}$, while for extremely high porosity ($f\lesssim0.01$) it is hard to produce scattering polarization.
Hence, moderate porosity dust particles have a role to relax the tight constraint on $a_\mathrm{max}$.

%

There is another issue concerning dust size constraints, that is, the inconsistency between opacity index and scattering polarization \citep{Kataoka16, Yang16a}. 
If the disk is optically thin at observing wavelengths, the spectral slope of the observed flux density depends on the opacity index $\beta$, where $\beta$ is the spectral slope of the absorption opacity, i.e., $\kappa_\mathrm{abs}\propto\lambda^{-\beta}$.
In Figure \ref{fig:beta}, the opacity index is shown as a function of $a_\mathrm{max}f$, where the opacity index is defined at wavelengths between $870~\mu$m and $1.3$ mm.
As shown in Figure \ref{fig:beta}, the opacity index at which scattering polarization can be anticipated is typically about $\beta\gtrsim1$. However, the opacity index of disks is typically equal to or less than unity \citep{Testi14}, implying the presence of mm to cm-sized dust particles in the disk \citep{Draine06}. Recently, \citet{Dent19} clearly showed that scattering polarization is detected at disk regions where $\beta\lesssim1$. This inconsistency also seems to occur for another disk \citep[e.g.,][]{Stephens17, Hull18}.

Dust porosity has been suggested as a solution for this inconsistency.
However, the opacity index of moderately porous particles is almost at the Rayleigh-limit value, that is, $\beta\approx\beta_\mathrm{ISM}\approx1.6$ \citep{Planck14}. 
Thus, this inconsistency may not be simply solved by considering particle porosity. Perhaps, other parameters, such as the functional shape of the dust size distribution and the optical depth of the disks, may be important to solve this issue. Recently, it has been pointed out that the spectral index could be affected by scattering if the disk is optically thick \citep{Liu19, Zhu19}. 

\subsubsection{Wavelength dependence}
\begin{figure}
\begin{center}
\includegraphics[height=6.0cm,keepaspectratio]{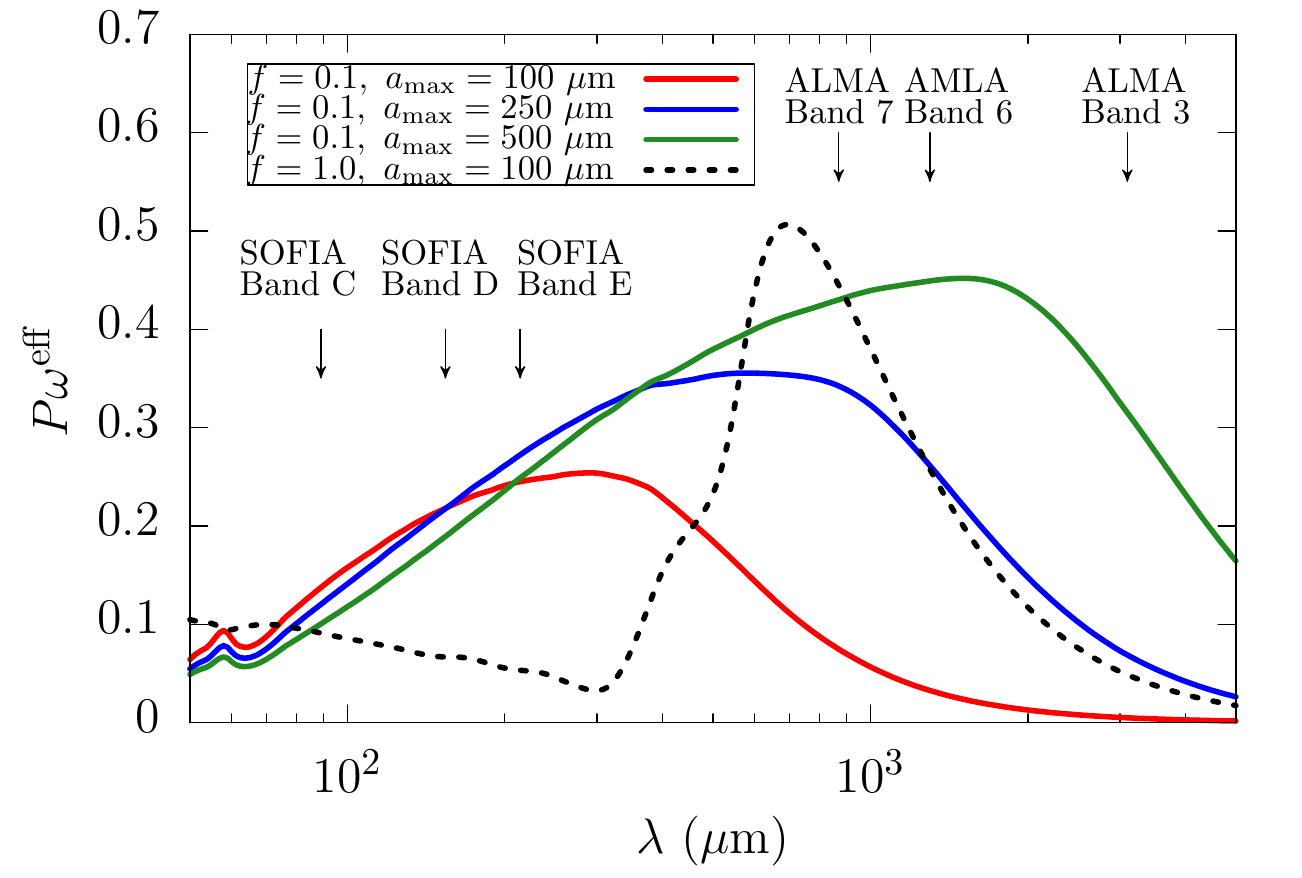}
\caption{Wavelength dependence of scattering polarization. Solid lines indicate compact aggregate models with $a_\mathrm{max}=100,~250,~500~\mu$m and $f=0.1$. Dashed line is the results for the solid sphere model with $a_\mathrm{max}=100~\mu$m. Wavelengths of some observational bands of the SOFIA telescope and ALMA are indicated by black arrows.}
\label{fig:lambdadep}
\end{center}
\end{figure}

The wavelength dependence of scattering polarization is another important point.
In Figure \ref{fig:lambdadep}, we show $P\omega^\mathrm{eff}$ as a function of wavelength.
For the case of $f=1$, scattering polarization efficiently occurs at $\lambda\approx2\pi a_\mathrm{max}$ \citep{Kataoka15}.
If dust porosity is taken into account, the wavelength dependence becomes weaker than that predicted from the solid sphere model as already expected from Figure \ref{fig:pomgeff_comp}.
Therefore, the wavelength dependence of scattering polarization is useful to constrain dust porosity. For example, if we detect scattering polarization both at Band 3 and 7 of ALMA, porous dust models seem to be favored as the origin of scattering polarization rather than the solid sphere model. 

\citet{Stephens17} found that the polarization pattern of HL Tau disk is wavelength dependent. At $\lambda=0.87$ mm, the polarization pattern of HL Tau's disk is consistent with a scattering origin, whereas at $\lambda=3$ mm, the polarization pattern becomes circular symmetric, indicating that another origin for the polarization is important, such as grain alignment \citep{Tazaki17, Kataoka17, Yang19}. DG Tau's disk also shows a wavelength dependent polarization pattern between $\lambda=0.87$ mm \citep{Bacciotti18} and $\lambda=3$ mm \citep{Harrison19}, where polarization at $\lambda=0.87$ mm is partially explained by scattering polarization and polarization at $\lambda=3$ mm is perhaps caused by grain alignment \citep{Harrison19}. 

The observed wavelength dependence of disks around HL Tau and DG Tau might be reproduced either with or without dust porosity (e.g., see the blue solid line and the dashed line in Figure \ref{fig:lambdadep}). 
Although wavelength dependences between ALMA bands is similar to each other, they have completely different wavelength dependence at far-infrared wavelengths. 
Therefore, far-infrared wavelength polarimetry, such as by the SOFIA telescope and also the future SPICA telescope, is essential to distinguish between these models.
The SOFIA and SPICA telescopes are insufficient to spatially resolve most disks, and hence, axisymmetric polarization patterns will be canceled out.
However, scattering polarization tends to have a unidirectional polarization pattern for inclined disks, and therefore, we may expect to detect polarized waves even if the polarization pattern is integrated over the entire disk. It is worth keeping in mind that at far-infrared wavelengths, disks may become optically thick. \citet{Yang17} pointed out that large optical depths of disks may reduce the scattering polarization fraction. 

\subsection{Implications for planetesimal formation} \label{sec:impl}
Porosity evolution of dust aggregates in disks has been simulated by a number of authors \citep{Ormel07, Okuzumi09, Zsom10, Okuzumi12, Kataoka13, Krijt15, Krijt16, Lorek18}. However, how dust aggregates grow to form planetesimals is a matter of debate. Here, we discuss implications for dust evolution and planetesimal formation by assuming that particles responsible for polarized scattered light dominate the entire population in disks. 

A possible scenario to form a planetesimal is direct coagulation via fluffy dust aggregates. Dust aggregates consisting of sub-micron-sized icy monomers are found to be very sticky, and they are resistant to compaction and fragmentation upon high-speed collisions \citep{Suyama08, Wada08, Wada09}. By assuming perfect sticking, fluffy dust aggregates ($f\approx10^{-4}$) form in disks and they finally become planetesimals by direct coagulation \citep{Okuzumi12, Kataoka13}. These results have also been confirmed by more recent studies \citep{Krijt15, Krijt16, Lorek18}.
However, as we mentioned in Section \ref{sec:constf}, these aggregates are not favored from observations of millimeter-wave scattering polarization. Although the presence of these aggregates cannot be ruled out only by scattering-polarization observations, the presence of relatively compact particles are not anticipated in this model. 

Therefore, a dust evolution model should answer how relatively compact particles could be formed in disks in order to explain millimeter-wave scattering polarization. 

The first possibility is dust compaction due to bouncing collisions. Laboratory experiments suggest that sequential bouncing collisions lead to gradual compaction of a dust aggregate \citep{Weidling09}. If bouncing collisions occur, \citet{Zsom10} suggested that the volume filling factor is increased up to 0.36 within $10^4$ orbital timescales, which is consistent with the scattering polarization constraints.
\citet{Windmark12a} show that dust coagulation stalls at about $100~\mu$m, which is also similar to the dust size expected from scattering polarization, although the dust size where coagulation stalls depends on disk properties, such as gas density, temperature, and turbulent strength. The dust evolution from a bouncing scenario seems to be consistent with that inferred from scattering polarization observations; however, the onset of bouncing collisions is still a matter of debate \citep{Wada11, Seizinger13a, Kothe13, Brisset17}, and hence, further studies are necessary to draw more robust conclusions. 

A second possibility is the increase of the internal density of dust aggregates caused by dust collisions with high-mass ratio \citep{Okuzumi09, Dominik16}. For example, monomer-aggregate collisions without restructuring produces aggregates with $d_f=3$ with $f\approx0.15$ which is known as BPCA. Such collisions are expected when fragmentation of dust aggregates occurs efficiently, and produces tiny fragments as small as monomer particle \citep{Wada08, Paszun09}. It was suggested that fragmentation of dust aggregates seems to be necessary to explain disk infrared observations \citep{Dullemond05}. 

In addition, recent laboratory measurements suggested that the adhesion energy of icy dust might be smaller than previous estimates, implying importance of dust fragmentation in disks. \citet{Musiolik16a, Musiolik16b} showed that CO$_2$-ice, which is also expected to condense onto dust particles in outer disk regions, does not show a high adhesion energy. Hence, dust aggregates of CO$_2$-ice-coated particles are thought to be much more fragile than those of H$_2$O-ice. It is worth mentioning that recent experiments also question a high adhesion energy of H$_2$O-ice at low temperature ($T\lesssim150-200$ K) \citep{Gundlach18, Musiolik19}, although earlier works by \citet{Gundlach15} confirmed high adhesion energy even at temperatures down to 100 K. 

Motivated by these laboratory experiments, \citet{Okuzumi19} have performed a dust coagulation simulation taking the non-sticky properties of the CO$_2$-ice-coated particles. 
Since lower adhesion energy predicts lower critical fragmentation velocity \citep{Dominik97}, the maximum dust radius is also reduced if the fragmentation limits coagulation \citep{Birnstiel11, Birnstiel12}. As a result, \citet{Okuzumi19} showed that scattering-like polarization of HL Tau's disk can be successfully explained by considering fragmentation of CO$_2$-ice-coated particles. 

Although laboratory experiments are still inconclusive, efficient dust fragmentation and subsequent dust collisions with high-mass ratio may potentially explain the origin of the relatively compact dust aggregates suggested by millimeter-wave scattering polarization.

A third possibility is that they are produced as fragments of differentiated planetesimals. If planetesimals are large enough to be molten, that is, differentiated, its fragments can be very compact particles ($f\approx1$). Melting events of icy dust particles, such as more gentle version of chondrule forming events, may also explain compact particles. However, in these case, it is necessary to explain why $a_\mathrm{max}$ is adjusted to $\lambda/2\pi$. 

%
%
Although the origin of compact (sub-)millimeter-size particles is an open question, these particles are expected to form planetesimals via either coagulation with mass transfer \citep{Windmark12a, Windmark12b} or the Streaming instability \citep{Youdin05, Johansen07, Bai10a, Bai10b}.
\citet{Windmark12a} shows that if dust coagulation stalls at sub-millimeter size and if a small amount of larger seed particles are present, i.e., centimeter size, the seed particles grow to form planetesimals via mass transfer. 
Meanwhile, if compact dust particles have $St>10^{-2}$, where $St$ is the Stokes number and describes dynamical coupling between a dust particle and gas, they are subjected to the Streaming instability \citep{Drazkowska14}. Thus, the gravitational collapse of dust clumps led by the streaming instability is also another feasible pathway toward planetesimals.

\subsection{Comparison with cometary dust in the Solar System} \label{sec:67p}
Comets in our Solar System are thought to be primitive objects, and hence, they are regarded as a living fossil of icy planetesimals. Cometary dust particles provide useful insights into how they form in the early solar nebula.

Recently, the {\it Rosetta} orbiter followed the comet 67P/Churyumov-Gerasimenko (hereafter 67P) and conducted {\it in-situ} measurements of cometary dust particles. 
Comet 67P is a km-sized comet, and it is considered to be a primordial rubble pile object \citep{Davidsson16} with bulk density of 0.533$\pm$0.006 g cm$^{-3}$ \citep{Patzold16}. Three instruments onboard {\it Rosetta} are dedicated to analyze dust particles : MIDAS (Micro-Imaging Dust Analysis System), GIADA (Grain Impact Analyser and Dust Accumulator), and COSIMA (Cometary Secondly Ion Mass Analyzer). 

These instruments revealed that cometary dust seems to have two different families in terms of morphology. MIDAS (Micro-Imaging Dust Analysis System) uses an the atomic force microscope and provides 3D tomographic images of cometary dust particles.
MIDAS found that cometary dust particles seem to have two morphological populations: compact aggregate and fluffy aggregates \citep{Bentley16, Mannel16, Mannel19}. The dust aggregates analyzed by the MIDAS have sizes from a few to 10 $\mu$m, and their subunit diameter is typically about from $0.1~\mu$m \citep{Mannel19} to $1~\mu$m \citep{Bentley16, Mannel16}. In addition, \citet{Bentley16} measured fractal dimensions of fluffy aggregates and found $d_f=1.7\pm0.1$. GIADA measures the cross-section, momentum, and mass of each dust particle and is mainly sensitive to (sub-)millimeter-sized dust particles. GIADA also measured two populations of dust particles \citep{Dellacorte15, Fulle15, Fulle16b, Fulle16a}. \citet{Fulle16b, Fulle16a} show that fluffy aggregates detected by GIADA seem to have fractal dimension $d_f=1.87$. Also, \citet{Fulle16a} derived the volume filling factor of compact dust particles as $f=0.48\pm0.08$. Measurements by COSIMA also support two morphological populations of dust particles \citep{Langevin16, Lasue19}. 

Fluffy dust aggregates ($d_f\lesssim2$) detected by {\it Rosetta} seem to be primordial aggregates formed in the early solar nebula. However, \citet{Fulle16a} estimated a mass fraction of fluffy aggregates contained in the nucleus of comet 67P and it is only about 0.015\%. Hence, comet 67P likely consists of compact dust aggregates with $f=0.48\pm0.08$. If compact aggregates with $f\approx0.5$ are distributed in disks, they are sufficient to produce millimeter-wave scattering in disks as we showed in Sections \ref{sec:optprop} and \ref{sec:result}. Hence, dust particles seen in scattering polarization might be building blocks of cometary objects. 

In addition to three dust analyzers, OSIRIS (Optical, Spectroscopic, and Infrared Remote Imaging System) also provide images of the nucleus as well as dust particles. Based on the OSIRIS images, the tensile strength of comet 67P is estimated \citep{Groussin15, Basilevsky16}. Recently, \citet{Tatsuuma19} studied the tensile strength of dust aggregates and found that the tensile strength of comet 67P is reproduced when the monomer radius is between $3.3-220~\mu$m. 
Hence, sub-millimeter-sized solid spheres are candidates to explain the measured tensile strength. 

To summarize, dust particles seen in scattering polarization might be precursors of cometary objects. 
However, it is worth keeping in mind that the present-day cometary dust particles are biased to more compact dust structure because cometary particles are the end product of a long series of compaction events. Also, dust outflow from a cometary coma may also selectively remove fluffy dust aggregates.

\section{Summary} \label{sec:sum}
We have studied how dust structure and porosity affect scattering polarization at millimeter wavelength because these quantities are important to understand how planetesimals form in protoplanetary disks. First of all, we have computed the optical properties of solid spheres and compact/fluffy dust aggregates at millimeter wavelengths, and then radiative transfer simulations were performed in order to assess their influence on millimeter-wave scattering polarization. 

Our primary findings are as follows:
\begin{enumerate}
\item The effective single scattering albedo $\omega^\mathrm{eff}$ of compact aggregates with $f=0.01$ and fluffy aggregates with $f\ll0.01$ is shown to be very small. As a result, dust particles with higher fractal dimension and/or lower porosity are more favorable to explain scattering polarization observations (see Section \ref{sec:optprop} and also Figure \ref{fig:pomg}). This is confirmed by performing 3D radiative transfer simulations in disks (Section \ref{sec:result}).

\item The polarization pattern of a disk containing moderately porous particles shows near-and-far side asymmetry. Polarization angles at the disk near side tend to show azimuthal directions, whereas those of the far side show the direction parallel to the minor axis (Figure \ref{fig:result_emt}).

\item Although a high porosity is not preferred, for moderately porous particles, the width of $P\omega^\mathrm{eff}$ becomes broad, indicating that scattering polarization can be expected for a wider range of $a_\mathrm{max}$. This may relax the tight constraints on $a_\mathrm{max}$ for the solid sphere model (Figure \ref{fig:beta}).

\item The wavelength dependence of scattering polarization becomes weaker for moderately porous particles compared to solid spheres (Figure \ref{fig:pomgeff_comp}). Thus, multi-wavelength polarimetric observations by ALMA as well as far-infrared disk polarimery seems to be useful to constrain dust porosity in disks (Figure \ref{fig:lambdadep}).

\item Detection of scattering polarization from a disk requires the presence of relatively compact dust aggregates. Aggregates with higher fractal dimension and lower porosity are favored. Although we cannot rule out the presence of fluffy aggregates, at least some amount of compact dust particles should be formed in disks (Section \ref{sec:constf}).

\item Dust particles, which can cause millimeter-wave scattering, are similar to those contained in the comet 67P. Thus, icy planetesimals, like comets, might be formed from these dust particles via either the streaming instability or mass transfer.

\end{enumerate}

\acknowledgments
R.T. would like to thank Cornelis P. Dullemond for making the \texttt{RADMC-3D} code public. R.T. was supported by a Research Fellowship for Young Scientists from the Japan Society for the Promotion of Science (JSPS) (JP17J02411). This work was also supported by JSPS KAKENHI Grant Numbers JP19H05068 (R.T.), JP17H01103 (H.T.), JP18K13590 and JP19H05088 (A.K.), and JP19K03926 and JP18H05438 (S.O.).

\software{RADMC-3D \citep{Dullemond12}}

\appendix

\section{Derivation of the upper bound on the effective scattering opacity of fluffy aggregates} \label{sec:appB}
The effective scattering opacity of dust aggregates can be written by
\begin{equation}
\kappa_\mathrm{sca}(1-g)=\frac{1}{m}\frac{2\pi}{k^2}\int_{-1}^1 (1-\mu)S_{11,\mathrm{agg}}(\mu)d\mu, \label{eq:def}
\end{equation}
where $m$ is the mass of the dust aggregate, $k$ is the wave number, $S_{11,\mathrm{agg}}$ is the (1,1) element of the scattering matrix of dust aggregates, and $\mu=\cos\theta$, where $\theta$ is the scattering angle. Using the single scattering assumption, a scattering matrix element of dust aggregates can be written by 
\begin{equation}
S_{11,\mathrm{agg}}(\mu)=N^2S_{11,\mathrm{mono}}(\mu)\mathcal{S}(q),
\end{equation}
where $N$ is the number of monomers, $\mathcal{S}(q)$ is the static structure factor and $q=2k\sin(\theta/2)$ is the magnitude of the scattering vector \citep{Tazaki16}.
When $qR_g\gg1$ and $d_f=2$, we can approximately decompose the scattering phase function of fluffy dust aggregates by the sum of coherent and incoherent contribution:
\begin{eqnarray}
S_{11,\mathrm{agg}}(\mu)&=&S_{11,\mathrm{agg}}^{\mathrm{coherent}}+S_{11,\mathrm{agg}}^{\mathrm{incoherent}}, \label{eq:coh}\\
S_{11,\mathrm{agg}}^{\mathrm{coherent}}&\simeq&N^2S_{11,\mathrm{mono}}(\mu)\delta(\mu-1),\\
S_{11,\mathrm{agg}}^{\mathrm{incoherent}}&\simeq&NS_{11,\mathrm{mono}}(\mu)[(qR_0)^{-2}+1],
\end{eqnarray}
where we have used Equation (29) of \citet{Tazaki16}.
Using Equations (\ref{eq:def} and \ref{eq:coh}) and $qR_0\ll1$, we obtain
\begin{equation}
\kappa_\mathrm{sca}(1-g)\simeq\frac{1}{m}\frac{2\pi N}{k^2}\int_{-1}^{1} (1-\mu)(qR_0)^{-2}S_{11,\mathrm{mono}}(\mu)d\mu,
\end{equation}
It is worth noting that the coherent component (forward scattering) does not contribute to the effective albedo.
As a result, we obtain
\begin{equation}
\kappa_\mathrm{sca}^\mathrm{eff}\equiv\kappa_\mathrm{sca}(1-g)=\frac{\kappa_{\mathrm{sca,mono}}}{2x_0^2}.
\end{equation}

\bibliography{cite}



\end{document}